\documentclass{article}

\begin{document}

\begin{center}
{\large {\bf PROPERTY VALUES}}
\vspace{0.1in}

{\em R. Delbourgo}

{\small \noindent School of Mathematics and Physics, 
        University of Tasmania\\
        Private Bag 37 GPO, Hobart, Australia 7001 \vspace{1in} }

\end{center}

\begin{abstract}
By ascribing a complex anticommuting variable $\zeta$ to each basic {\em 
property} of a field it is possible to describe all the fundamental particles 
as combinations of only five $\zeta$ and understand the 
occurrence of particle generations. An extension of space-time $x$ to include 
property then specifies the `where, when and what' of an event and allows
for a generalized relativity where the gauge fields lie in the $x-\zeta$
sector and the Higgs fields in the $\zeta-\zeta$ sector.
\end{abstract}



\section*{Preamble}
First of all let me say why I am so pleased to be present at this FestSchrift 
in honour of Girish' and Bruce's retirements. When I arrived in Tasmania one 
of my first tasks was to make contact with physicists in other Australian 
institutions. Amongst these Melbourne University had high priority and I was 
very glad to welcome Bruce as one of the earliest visitors to Hobart. Since then 
we have interacted many times and I have relished my own visits to Melbourne
to give occasional seminars and find out what Joshi and McKellar were up to. 
Let me wish them both a long and happy retirement and say how much I have 
appreciated their support and friendship throughout the last 30 years! 
Retirement has much to recommend it and they might care to view their future 
condition as changing from `battery hen' to `free-range chicken', since they 
are no longer obliged to feed on grants in order to lay eggs.

\section{Introduction}
As most of you will have already surmised, the title of my talk has nothing
to do with real estate and perhaps sounds all the more mysterious for it. In fact
property values have everything do with quantum fields and accurately reflect 
the contents of what follows. The motivation for this work is to be able to
describe the `what' as well as the `where-when' of an event. To make an analogy
with personality traits in humans, psychologists may characterise a person as
optimistic/pessimistic, happy/sad, aggressive/submissive, etc. (A person will
possess some combination or superposition over these trait states.) The same sort
of characterisation applies to a quantum field and is usually underlined by 
attaching a label to each distinct field. Thus we speak of an electron/positron 
field, a neutron/antineutron field, a quark/antiquark field, etc. and draw them
in Feynman diagrams say with solid or dashed lines and arrows, plus legends if
necessary, to distinguish them from one another. An event is some confluence
between these labels with a possible interchange of traits, as specified by
an interaction Lagrangian. Often we assume a symmetry group is operational 
which `rotates' labels; this constrains the interchange of property and, if
the symmetry is local, it can be gauged.

The basic idea I wish to put to you is that traits/labels are normally discrete:
thus a field is either an electron or it is not; similarly for a proton, and so 
on. It makes sense to attach a separate coordinate to each such property and
choose the coordinate to be anticommuting since its occupation number is either
one or zero. (So a person can be pessimistic \& sad \& aggressive with a product
of the three traits.) Furthermore if we make the coordinate $\zeta$ complex we 
can describe the converse property by simple conjugation. When applying this
idea to quantum fields the question arises: how many coordinates are needed?

The fewer the better of course. From my investigations thus far \cite{RD1,RD2}
I have concluded that one can get a reasonable description of the known 
fundamental particle spectrum by using just five complex cordinates $\zeta^\mu;
\, \mu = 0,1,2,3,4$. I have found \cite{RD1} that four $\zeta$ are insufficient 
to account for the three known light generations and their features. The way 
generations arise in this 
context comes by multiplying traits by neutral products of other traits. 
For instance a person can be pessimistic or (pessimistic $\times$ sad-happy) or
(pessimistic $\times$ aggressive-submissive) and so on. With $N$ $\,\,\zeta$ one
potentially encounters $2^{N-1}$ generations of a particular property by forming
such neutral products; this is interesting because it suggests that any 
anticommuting property scheme produces an even number of generations. This is 
not a cause for panic; all we know at present is that there exist 3 light 
neutrinos, so there may be other (possibly sterile) heavy neutrinos accompanied
by other quarks and charged leptons.

The proposal therefore is to append a set of five complex anticommuting
coordinates $\zeta^\mu$ to space-time $x^m$ which are to be associated with
property. This is in contrast to traditional brane-string schemes which append yet 
unseen bosonic extra degrees of freedom to space-time. Now one of the most 
interesting aspects of fermionic degrees of freedom is that they act oppositely to 
bosonic ones in several respects: we are familiar with sign changes in commutation
relations, statistical formulae and, most importantly, quantum loop contributions; 
but less well-known is that in certain group theoretical representations \cite{Neg}
the $SO(2N)$ Casimirs are continuations to negative $N$ of $Sp(2N)$ Casimirs so 
that anticommuting coordinates {\em effectively subtract} dimensions. The
opposite (to bosons) loop sign is extra confirmation of this statement and 
this is put to great use in standard supersymmetry (SUSY) in order to resolve the 
fine-tuning problem. So it is not inconceivable that with a correct set of
anticommuting property coordinates $\zeta$ appended to $x$ we might end up
with zero net dimensions as it was presumably before the BIG BANG; at the very 
least that we may arrive at a scheme where fermions cancel quantum effects from 
bosons as in SUSY. For this it is not necessary to embrace all the tenets of 
standard SUSY; indeed we shall contemplate an edifice where property coordinates 
are Lorentz scalar. In that respect they are similar to the variables occurring 
in BRST transformations for quantized vector gauge-fixed models --- without 
implying any violation of the spin-statistics theorem for normal physical fields.

\section{Property superfields}
Now let us get down to the nuts and bolts of the construction and find out what
the edifice will look like. We might be tempted to match the four space-time $x^m$ 
by four $\zeta^\mu$, but as we have shown elsewhere that is not enough -- lepton
generations do not ensue. However, all is not lost: we can add 5 $\zeta^\mu$
(which are sufficient), but only take half the states to get correct statistics. 
Associate each Lorentz scalar anticommuting numbers with a `property' or `trait';
this invites us to consider symmetry groups\cite{Groups} like $SU(5)$ or $Sp(10)$ 
or $SO(10)$ to reshuffle the properties. (That these popular groups pop up is 
probably no accident.) The only way I have found to obtain the well-established 
quantum numbers of the fundamental particle spectrum as superpositions over traits
is to postulate the following charge $Q$ and fermion number $F$ assignments,
\[ 
Q(\zeta^{0,1,2,3,4}) = (0,1/3,1/3,1/3,-1);\quad
F(\zeta^{0,1,2,3,4}) = (1,-1/3,-1/3,-1/3,1). 
\]
Other properties are to built up as composites of these. For example $\zeta^4$
may be identified as a negatively charged lepton, but then so can its product 
with neutral combinations like $\bar{\zeta}_0\zeta^0$, $\bar{\zeta}_i\zeta^i$,...
strongly suggesting how generations can arise in this framework. We'll
return to this shortly.

Since the product of an even number of $\zeta$ is a (nilpotent) commuting property 
a Bose superfield $\Phi$ should be a Taylor series in even powers of 
$\zeta,\bar{\zeta}$. Similarly a superfield which encompasses fermions
$\Psi_\alpha$ will be a series in odd powers of $\zeta,\bar{\zeta}$ --- 
up to the 5th:
\[\Phi(x,\zeta,\bar{\zeta})=\sum_{even\,r+\bar{r}}
 (\bar{\zeta})^{\bar{r}}\phi_{(\bar{r}),(r)}(\zeta)^r;
\]
\[
 \Psi_\alpha(x,\zeta,\bar{\zeta})= \sum_{odd\, r+\bar{r}}
 (\bar{\zeta})^{\bar{r}}\psi_{\alpha(\bar{r}),(r)}(\zeta)^r.
\]
So far as labels $\mu$ on $\zeta^\mu$ go, we can characterise
\begin{itemize}
\item label 0 as {neutrinicity}
\item labels {1 - 3} as (down) {chromicity}
\item label 4 as {charged leptonicity}
\end{itemize}
\newpage

You will notice that these expansions produce too many states for $\psi_\alpha$ 
and $\phi$, viz. 512 properties in all, so they demand cutting down. 
An obvious tactic is to associate conjugation ($^c$) with the operation $\zeta 
\leftrightarrow \bar{\zeta}$
\[
\psi_{(r),(\bar{r})} = \psi^{(c)}_{(\bar{r}),(r)},
\]
corresponding to reflection along the main diagonal when we expand
$\Psi$ as per the table below. Indeed if we suppose that
all fermion field components are left-handed the conjugation/reflection operation
then automatically includes right-handed particle states. Even so there remain
too many components and we may wish to prune more. One strategy is to notice
that under reflection about the {\em cross-diagonal} the $F$ and $Q$ quantum 
numbers are not altered. We shall call this cross-diagonal reflection a 
{\em duality} ($^\times$) transformation. For example,
\[
(\bar{\zeta}_\alpha\zeta^\mu\zeta^\nu)^{\times} = \frac{1}{3!}
 \epsilon^{\rho\sigma\tau\mu\nu}\bar{\zeta}_\rho\bar{\zeta}_\sigma
  \bar{\zeta}_\tau.\frac{1}{4!}\epsilon_{\alpha\beta\gamma\delta\epsilon}
                     \zeta^\beta\zeta^\gamma\zeta^\delta\zeta^\epsilon.
\]
By imposing the antidual reflection symmetry: $\psi_{(r),(\bar{r})} = 
- \psi_{(5-\bar{r}),(5-r)}$ we roughly halve the remaining number of 
components. So whereas previously we had the separate set of neutrino states,
for instance,
\[
\zeta^0,\zeta^0(\bar{\zeta}_4\zeta^4),\zeta^0(\bar{\zeta}_i\zeta^i),
\zeta^0(\bar{\zeta}_4\zeta^4)(\bar{\zeta}_i\zeta^i),
\zeta^0(\bar{\zeta}_i\zeta^i)^2,\,(\bar{\zeta}_i\zeta^i)^3,
\zeta^0(\bar{\zeta}_4\zeta^4)(\bar{\zeta}_i\zeta^i)^2,
\zeta^0(\bar{\zeta}_4\zeta^4)(\bar{\zeta}_i\zeta^i)^3
\]
antiduality sifts out half the combinations, namely:
\[
\zeta^0[1-(\bar{\zeta}_4\zeta^4)(\bar{\zeta}_i\zeta^i)^3/6],\quad
\zeta^0[\bar{\zeta}_4\zeta^4)-(\bar{\zeta}_i\zeta^i)^2/2],
\]
\[
\zeta^0[(\bar{\zeta}_i\zeta^i)-(\bar{\zeta}_4\zeta^4)(\bar{\zeta}_j\zeta^j)^2/2],
\quad\zeta^0[(\bar{\zeta}_i\zeta^i)(\bar{\zeta}_4\zeta^4)-(\bar{\zeta}_j\zeta^j)^2/2]
\]
In particular as $\bar{\zeta}_0\bar{\zeta}_4\zeta^1\zeta^2\zeta^3$ and 
$\bar{\zeta}_4\zeta^0\zeta^1\zeta^2\zeta^3$ are self-dual, imposing 
antiduality eliminates these unwanted states, a good thing since they respectively 
have $F=3$ and $Q=-2$. Applying antiduality and focussing on left-$\Psi$, 
the resulting square contains the following varieties of up ($U$), down ($D$), 
charged lepton ($L$) and neutrinos ($N$), where the subscript distinguishes 
between repetitions. In the $\zeta\bar{\zeta}$ expansion table,
$\times$ are duals, * are conjugates:
\begin{center}
 \begin{tabular}{|l||c|c|c|c|c|c|}  \hline
  $r\backslash\bar{r}$ & 0 & 1 & 2 & 3 & 4 & 5 \\
  \hline \hline
   0 &   & $\!\!L_1,N_1,D_5^c$ &  & $L_5^c,D_1,U_1$ & &   \\
   1&*&&$\!\!L_{2,3},N_{2,3},D^c_{3,6,7},U_3^c$
             & &$\!\!L_6^c,D_2,U_2$&\\
   2 &   &     *     &          & $\!\!L_4,N_4,D^c_{4,8},U_4^c$&&$\times$ \\
   3 & * &           &    *     &        &  $\times$   &   \\
   4 &   &     *     &          &   *    &      & $\times$ \\
   5 & * &           &    *     &        &   *  &    \\
 \hline 
 \end{tabular}
\end{center}

Observe that colour singlet and triplet fermions come in 4s, 6s and 8s which 
comfortably contain the
known three generations. However you will see that whereas $U,D$ in the first and 
second generation are bona fide weak isospin doublets, the third and fourth
family $U,D$ are accompanied by another exotic colour triplet quark, call it $X$ 
say, having charge $Q =-4/3$, and make up a weak isospin triplet.
So this is a departure from the standard model! Specifically the weak isospin 
generators are:
\[
T_+ = \zeta^0\partial_4 - \bar{\zeta}_4\bar{\partial}^0, \quad
  T_- = \zeta^4\partial_0 - \bar{\zeta}_0\bar{\partial}^4; 
\]
\[
2T_3 = [T_+,T_-] = \zeta^0\partial_0-\zeta^4\partial_4+
     \bar{\zeta}_4\bar{\partial}^4-\bar{\zeta}_0\bar{\partial}^0,\,
{\rm so~we~meet}
\]
\[
{\rm doublets~like~}  (N_1,L_1), (U_1,D_1)\sim (\zeta^0,\zeta^4),
    \quad (-\bar{\zeta}_4,\bar{\zeta}_0)
\]
\[
{\rm singlets~like~}L_5\,\,D_5 \sim 
   (\bar{\zeta}_0\zeta^0 + \bar{\zeta}_4\zeta^4),
  \quad (\bar{\zeta}_0\zeta^0\bar{\zeta}_4\zeta^4),
\]
\[
{\rm triplets~like~} (U_3,D_3,X_3)\sim (-\bar{\zeta}_4\zeta^0,
   [\bar{\zeta}_0\zeta^0-\bar{\zeta}_4\zeta^4]/\sqrt{2},\bar{\zeta}_0\zeta^4).
\]

\section{Exotic Particles, Generations \& the Mass Matrix}
The $U$-states arise from combinations like $\zeta^i\zeta^j\zeta^0$ lying in the
$\overline{10}$-fold SU(5) combination $\zeta^\lambda\zeta^\mu\zeta^\nu$ and as 
$\bar{\zeta}_k\bar{\zeta}_4\zeta^0$. (Note that $\zeta^0\zeta^4\zeta^k$ has the 
exotic value $F=5/3$ and cannot be identified with $U^c$.) The $D$-states occur 
similarly, as do the $N$'s and $L$'s. However, observe that $L_{5,6}\sim 
\bar{\zeta}_3\bar{\zeta}_2\bar{\zeta}_1$ and $D_{5,6,7,8} \sim \bar{\zeta}_k$
are nominally weak isosinglets, which again {\em differs} from the standard model. 
These mysterious states are definitely charged but do not possess any weak
interactions, so their behaviour is very curious. One might even regard $L_{5,6}$ 
like colourless charged baryons (antiproton-like) but really until one
sees how these states mix with the usual weak isodoublet leptons it is
dangerous to label them one thing rather than another without further research.

`Pentaquarks' such as $\Theta^+ \sim uudd\bar{s}$ \& $\Xi^{--} \sim ddss\bar{u}$ 
were recently discovered(?) with quite narrow widths and many people have 
advanced models to describe these new resonances as well as tetraquark mesons. 
But who is to say unequivocally that they are not composites of ordinary 
quarks and another $U$-quark or other $D$-family quarks which my scheme indicates? 
Further, I get a fourth neutrino, which could be essentially sterile and 
might help explain the mystery of $\nu$ masses. This is clearly fertile 
ground for investigation and I have only scratched the surface here.
Possibly conflicts with experiment may arise that will ultimately
invalidate the entire property values scheme.

One of the first matters to be cleared up is the mass matrix and flavour 
mixing which affects quarks as well as leptons. If we assume it is due to a 
Higgs $\Phi$ field's expectation values, there are nine colourless possibilities 
having $F=Q=0$ lying in an antidual boson superfield:
\begin{itemize}
 \item one $\phi_{(0)(0)} = \langle\phi\rangle = {\it M}$ 
 \item one $\phi_{(0)(4)} = \langle\phi_{1234}\rangle 
 = {\it H}{\rm ~complex}$
 \item three $\phi_{(1)(1)}=\langle\phi^0_0,\phi_4^4,\phi_i^i\rangle 
 = {\it A,B,C}$
 \item four $\phi_{(2)(2)}=\langle\phi_{04}^{04},\phi_{0k}^{0k},
  \phi_{4k}^{4k},\phi_{ij}^{ij}\rangle = {\it D,E,F,G},$ 
\end{itemize}
others being related by duality. With nine $\langle\phi\rangle$ the mass matrix 
calculation already becomes a difficult task! To ascertain what happens, 
consider the subset of quarks
involving $U_{1-4}$ and $D_{1-4,7}$ interacting with anti-selfdual superHiggs.

The Lagrangian $\int d^5\zeta d^5\bar{\zeta}\,\,\bar{\Psi}\langle\Phi\rangle\Psi$
produces $U$ and simplistic $D$ mixing matrices:
\[
 2M(U)\!\rightarrow\!\left( \begin{array}{cccc}
   2{\it M}+{\it F}/\sqrt{3}&{\it B}+{\it C}/\sqrt{3}&-{\it H}^* & 0\\
   {\it B}+{\it C}/\sqrt{3} & 2{\it M} & 0 & 0 \\
   -{\it H} & 0 & 2{\it M}+{\it G}/\sqrt{3} & 2{\it C}/\sqrt{3} \\
   0 & 0  & 2{\it C}/\sqrt{3} & 2{\it M}
 \end{array} \right)
\]
\[2M(D)\!\!\rightarrow\!\!\left( \begin{array}{ccccc}
       \!\!2{\it M}+{\it E}/\sqrt{3}&\!{\it A}+{\it C}/\sqrt{3}& -{\it H}^*&0&0\\
       \!\!{\it A}+{\it C}/\sqrt{3} & 2{\it M}& 0 & 0 & 0 \\
       -{\it H}&0& 2{\it M}\!+{\it G}/\sqrt{3}&2{\it C}/\sqrt{3}&\!
       ({\it E}\!-\!{\it F})/\sqrt{3}\\       
       0&0& 2{\it C}/\sqrt{3} & 2{\it M} & {\it A}-{\it B} \\
       0 & 0 & -({\it E}\!+\!{\it F})/{\sqrt{3}} & {\it A}-{\it B}
       & 2{\it M}-{\it D} \\
 \end{array} \right).
\]
Similar expressions can be found for the leptons and quarks.
Such matrices have to be diagonalised via unitary transformations $V(U)$ \& $V(D)$.
However the coupling of the weak bosons is {\em different} for 
isodoublets $U_{1,2}~D_{1,2}$ and isotriplets $U_{3,4}~D_{3,4}$ (and zero for the 
weak isosinglet $D_7$). So we can't just evaluate $V^{-1}(U)V(D)$ for the unitary
CKM matrix now. Rather the weak interactions connecting $U$ \& $D$ 
mass-diagonalised quarks will not be quite unitary (because of the different 
coupling factors):
\[
L_{\rm weak}/g_w = W^+(\bar{U}_1D_1 +\bar{U}_2D_2 )
       +\sqrt{2}W^+(\bar{U}_3D_3+\bar{U}_4D_4+\bar{D_3}X_3+\bar{D}_4X_4)
\]
\[
 \quad\quad + W^-(\bar{D}_1U_1 +\bar{D}_2U_2 )
       +\sqrt{2}W^-(\bar{D}_3U_3+\bar{D}_4U_4+\bar{X_3}D_3+\bar{X}_4D_4)+ 
    W^3{\rm ~terms}
\] 
So this is another departure from the standard picture: nonunitarity of the 
$3\times 3$ CKM matrix is a test of the scheme. Also CP violation is an 
intrinsic feature of the property formalism because $\it H$ is naturally complex, 
unlike the other expectation values ${\it A,B...M}$. A more realistic attempt
for getting the quark and lepton masses would be to abandon antiduality in
the Higgs sector and use all 18 expectation values; otherwise it may prove
impossible to cover the 12 or more orders of magnitude all the way from the
electron neutrino to the top quark (and higher).

\section{Generalized relativity}
We know that gauge fields can transport/communicate property from one 
place to another so where are they? Maybe one can mimic the SUSY procedure
and supergauge the massless free action for $\Psi$, without added complication
of spin. But there is a more compelling way, which has the benefit of 
incorporating gravity. Construct a fermionic version of Kaluza-Klein (KK) 
theory \cite{KKF}, this time without worrying about infinite modes which arise 
from squeezing normal bosonic coordinates. These are the significant points
of such an approach:
\begin{itemize}
 \item One must introduce a fundamental length $\Lambda$ in the extended $X$, as
   property $\zeta$ has no dimensions; maybe this is the gravity scale $\kappa=
  \sqrt{8\pi G_N}$?
 \item Gravity (plus gauge field products) fall within the $x-x$ sector,
  gauge fields in $x-\zeta$ and the Higgs scalars must form a matrix in
   $\zeta-\zeta$, 
 \item Gauge invariance is connected with the number of $\zeta$ so $SU(5)$ or 
 $Sp(10)$ {\em or perhaps a subgroup} are indicated,
 \item There is no place for a gravitino as spin is absent ($\zeta$ are
   Lorentz scalar),
 \item There are necessarily a small finite number of modes,
 \item Weak left-handed SU(2) is associated with rotations of $\zeta$ {\bf not}
  $\bar{\zeta}$ so may have something to with $\zeta$-analyticity.
\end{itemize}

The real metric specifies the separation in location as well as property:
it tells us how `far apart' and `different in type' two events are. Setting
$\bar{\zeta}^{\bar{\mu}}\equiv \bar{\zeta}_\mu$,
\[
ds^2 = dx^mdx^n{G_{nm}} + dx^md\zeta^\nu{G_{\nu m}} 
    + dx^n d\bar{\zeta}^{\bar{\mu}}{G_{\bar{\mu} n}} + 
  d\bar{\zeta}^{\bar{\mu}}d\zeta^\nu{G_{\bar{\mu}\nu}}
\]
where the tangent space limit corresponds to Minkowskian 
\[
 G_{ab}\rightarrow I_{ab}= \eta_{ab},\, {G_{\bar{\alpha}\beta}}\rightarrow
 I_{\bar{\alpha}\beta} = \Lambda^2{\delta_{\alpha\beta}},
\]
multiplied at least by $(\bar{\zeta}\zeta)^5$ --- to arrange correct property 
integration. Proceeding to curved space the components should contain the force 
fields, leading one to a `superbein'
\[
 {\bar{E}_M}^A = \left( \begin{array}{cc}{{e_m}^a}&
 {i\Lambda(A_m)_\mu^\alpha\zeta^\mu} \\
    0 & {\Lambda\delta_\mu^\alpha} \end{array} \right),
\]
and the metric ``tensor'' arising from
\[ds^2 = dx^m dx^n g_{nm}+2\Lambda^2
     [d\bar{\zeta}^{\bar{\mu}}-idx^m\bar{\zeta}^{\bar{\kappa}}
      (A_m)_{\bar{\kappa}}^{\bar{\mu}}]\delta_{{\bar{\mu}}\nu}
     [d\zeta^\nu + idx^n (A_n)^\nu_\lambda\zeta^\lambda];
\]
\[
 g_{mn} = e_m^ae_n^b\eta_{ab}.
\]

Gauge symmetry corresponds to the special change $\zeta^\mu\rightarrow
\zeta'^\mu=[\exp(i\Theta(x))]^\mu_\nu\zeta^\nu $ with $x'=x$. Given the standard 
transformation law
\[
{G_{\zeta m}}(X)=\frac{\partial X^{\prime R}}{\partial x^m}
         \frac{\partial X^{\prime S}}{\partial\zeta} G'_{SR}(X')(-1)^{[R]}
 ={\frac{\partial \zeta'}{\partial\zeta}}{G'_{\zeta m}}-
   {\frac{\partial \bar{\zeta}'}{\partial x^m}}{
    \frac{\partial \zeta'}{\partial\zeta}G'_{\zeta\bar{\zeta}}}.
\]
this translates into the usual gauge variation (a matrix in property space),
\[
A_m(x) =\exp(-i\Theta(x))[A'_m(x) - i\partial_m]\exp(i\Theta(x)).
\]
The result is consistent with other components of the metric tensor but
does not fix what (sub)group is to be gauged in property space
although one most certainly expects to take in the nonabelian colour group and 
the abelian electromagnetic group, so as to agree with physics.

\section{Some notational niceties}
When dealing with commuting and anticommuting numbers within a single 
coordinate framework
$X^M = (x^m,\zeta^\mu)$ 
one has to be {\bf exceedingly} careful with the order of quantities and of
labels. I cannot stress this enough. (It took me six months and much heartache
to get the formulae below correct.) For derivatives the rule is 
$dF(X) = dX^M (\partial F/\partial X^M) 
\equiv dX^M \partial_M F$, {\bf not} with the $dX$ on the right, and for 
products of functions: $d(FG..)=dF\,G+F\,dG+..$.  Coordinate transformations 
read $dX^{\prime M}=dX^N(\partial X^{\prime M}/\partial X^N)$, 
and {\em in that particular order}. Since $ds^2 = dX^N dX^M \,G_{MN}$, the
symmetry property of the metric is $G_{MN}= (-1)^{[M][N]}G_{NM}$.
Let $G^{LM}G_{MN} = \delta_N^L$ for the inverse metric, so
$G^{MN}=(-1)^{[M]+[N]+[M][N]}\,G^{NM}$. The notation here is $[M]=0$ for
bosons and 1 for fermions.

Changing coordinate system from $X$ to $X'$, we have to be punctilious 
with signs and orders of products, things we normally never care
about; the correct transformation law is
$$G_{NM}(X) = \left(\frac{\partial X^{\prime R}}{\partial X^M}\right)\left(\frac
 {\partial X^S}{\partial X^N}\right)G'_{SR}(X')\,(-1)^{[N]([R]+[M])}.$$
Transformation laws for contravariant and covariant vectors read:
$$ V^{\prime M}(X') = V^R(X)\left(\frac{\partial X^{\prime M}}{\partial X^R}\right) \quad
{\rm and}\quad A'_M(X')=\left(\frac{\partial X^R}
{\partial X^{\prime M}}\right)A_R(X),$$
in the order stated. Thus the invariant contraction is
$$V^{\prime M}(X')A'_M(X') = V^R(X)A_R(X) = (-1)^{[R]}A_R(X)V^R(X).$$ The inverse
metric $G^{MN}$ can be used to raise and lower indices as well as forming 
invariants, so for instance $V_R\equiv  V^SG_{SR}$ and $V^{\prime R}V^{\prime S}
 G'_{SR} = V^MV^NG_{NM}$.

The next issue is covariant differentiation; we insist that $A_{M;N}$ should
transform like $T_{MN}$, viz.
$$T'_{MN}(X') =(-1)^{[S]+[N])[R]}\left(\frac{\partial X^R}{\partial X^{\prime M}}
 \right)\left(\frac{\partial X^S}{\partial X^{\prime N}}\right) T_{RS}(X).$$
After some work we find that
$$A_{M;N} = (-1)^{[M][N]}A_{M,N} - A^L\Gamma_{\{MN,L\}},$$
where the connection is given by
\begin{eqnarray*}
\Gamma_{\{MN,L\}}&\equiv& [(-1)^{([L]+[M])[N]}G_{LM,N} +(-1)^{[M][L]}G_{LN,M}
     -G_{MN,L}]/2 \\
 &=& (-1)^{[M][N]}\Gamma_{\{NM,L\}}.\nonumber
\end{eqnarray*}
Another useful formula is the raised connection
\[
{\Gamma_{MN}}^K \equiv (-1)^{[L]([M]+[N])}\Gamma_{\{MN,L\}}G^{LK} 
  = (-1)^{[M][N]}{\Gamma_{NM}}^K,
\]
whereupon one may write
\[A_{M;N} = (-1)^{[M][N]}A_{M,N} - {\Gamma_{MN}}^LA_L.
\]
Similarly one can show that for double index tensors the true differentiation
rule is
\[
T_{LM;N}\!\equiv\!(-1)^{[N]([L]\!+\![M])}T_{LM,N}-
 (-1)^{[M][N]}{\Gamma_{LN}}^KT_{KM}
  -(-1)^{[L]([M]\!+\![N]\!+\![K])}{\Gamma_{MN}}^KT_{LK}.
\]
As a nice check, the covariant derivative of the metric properly vanishes:
$$G_{LM;N} \equiv (-1)^{[N]([L]+[M])}G_{LM,N} -(-1)^{[L][M]}{\Gamma}_{\{LN,M\}}-
   \Gamma_{\{MN,L\}} \equiv 0.$$

Moving on to the Riemann curvature we form doubly covariant derivatives:
$$A_{K;L;M} - (-1)^{[L][M]}A_{K;M;L} \equiv (-1)^{[K]([L]+[M])}{R^J}_{KLM}A_J$$
where one discovers that
\begin{eqnarray*}
 {R^J}_{KLM}&\equiv& (-1)^{[K][M]}({\Gamma_{KM}}^J)_{,L} 
                   - (-1)^{[L]([K]+[M])}({\Gamma_{KL}}^J)_{,M} \nonumber \\
      &  &   + (-1)^{[M]([K]+[L])+[K][L]}{\Gamma_{KM}}^N{\Gamma_{NL}}^J
                   - (-1)^{[K]([M]+[L])}{\Gamma_{KL}}^N{\Gamma_{NM}}^J.
\end{eqnarray*}
Evidently, ${R^J}_{KLM} = -(-1)^{[L][M]}{R^J}_{KML}$ and, less obviously, the
cyclical relation takes the form 
\[
(-1)^{[K][L]}{R^J}_{KLM} + 
(-1)^{[L][M]}{R^J}_{LMK}+ (-1)^{[M][K]}{R^J}_{MKL}=0.
\]
The fully covariant Riemann tensor is
$R_{JKLM}\equiv (-1)^{([J]+[K])[L]}{R^N}_{KLM}G_{NJ}$ with pleasing features:
\begin{eqnarray*}
R_{JKLM} & = & -(-1)^{[L][M]}R_{JKML} = -(-1)^{[J][K]}R_{KJLM},\\
0 & = & (-1)^{[J][L]}R_{JKLM}+(-1)^{[J][M]}R_{JLMK}+(-1)^{[J][K]}R_{JMKL} \\
R_{JKLM}& = &  (-1)^{([J]+[K])([L]+[M])} R_{LMJK}.\nonumber
\end{eqnarray*}
Finally proceed to the Ricci tensor and scalar curvature:
\begin{eqnarray*}
R_{KM} &\equiv& (-1)^{[J]+[K][L]+[J]([K]+[M])}G^{LJ}R_{JKLM}\nonumber \\
& = &(-1)^{[L]([K]+[L]+[M])}{R^L}_{KLM}\!=\!(-1)^{[K][M]}R_{MK},
\end{eqnarray*}
\[ R\equiv  G^{MK}R_{KM}.\]

\section{Curvatures of space-time-property}
When Einstein produced his general theory of relativity with Grossmann he wrote
all his expressions in terms of real variables. In order to avoid any confusion
with complex variables, we shall copy him by writing everything in terms of real 
coordinates $\xi,\eta$ rather than complex $\zeta=(\xi+i\eta)/\sqrt{2}$. 
(Note that the real invariant is $\bar{\zeta}\zeta = i\xi\eta$ and that a phase
transformation of $\zeta$ corresponds to a real rotation in $(\xi,\eta)$ space.)
For simplicity consider just one extra pair (rather than five pairs) and 
the following two examples, which are complicated enough as it is. 

\newpage
\noindent
(1) Decoupled property and space-time, but both curved:
\begin{eqnarray*}
ds^2&=&dx^mdx^nG_{nm}(x,\xi,\eta)+2id\xi d\eta G_{\eta\xi}(x,\xi,\eta)\nonumber
 \\
 &\equiv& dx^mdx^ng_{nm}(x)(1+if\xi\eta)+2i\Lambda^2d\xi d\eta(1+ig\xi\eta)
\end{eqnarray*}
from which we can read off the metric components ($G^{LM}G_{MN}\equiv\delta^L_N$)
\[
  \left( \begin{array}{ccc}
   G_{mn} & G_{m\xi} & G_{m\eta}\\
   G_{\xi n} & {G_{\xi\xi}} & {G_{\xi\eta}}\\
   G_{\eta n} & {G_{\eta\xi}} & {G_{\eta\eta}}
          \end{array} \right)  =
 \left( \begin{array}{ccc}
           g_{mn}(1+if\xi\eta) & 0 & 0 \\
           0  & 0 & -i\Lambda^2(1+ig\xi\eta)\\
           0 & i\Lambda^2(1+ig\xi\eta) & 0
          \end{array} \right), 
\]
\[
  \left( \begin{array}{ccc}
   G^{lm} & G^{l\xi} & G^{l\eta}\\
   G^{\xi m} & {G^{\xi\xi}} & G^{\xi\eta}\\
   G^{\eta m} & G^{\eta\xi} & G^{\eta\eta}
          \end{array} \right)  =
 \left( \begin{array}{ccc}
           g^{lm}(1-if\xi\eta) & 0 & 0 \\
           0  & 0 & -i(1-ig\xi\eta)/\Lambda^2\\
           0 & i(1-ig\xi\eta)/\Lambda^2 & 0
          \end{array} \right).
\]
The non-zero connections in the property sector are
\[
 \Gamma_{\xi\eta}^\xi = -{\Gamma_{\eta\xi}}^\xi = ig\xi,\quad
  {\Gamma_{\xi\eta}}^\eta = -{\Gamma_{\eta\xi}}^\eta = ig\eta 
\]
so
\begin{eqnarray*}{
{R^\eta}_{\xi\eta\eta}}&=& -2ig(1+ig\xi\eta) = 
-{{R^\xi}_{\eta\xi\xi}}\\
{{R^\xi}_{\xi\xi\eta}}&=& -ig(1+ig\xi\eta) =-
{{R^\eta}_{\eta\eta\xi}}\\
{\rm so~}{R_{\xi\eta}}&=&-{R_{\eta\xi}}=3ig(1+ig\xi\eta).
\end{eqnarray*}
Consequently the total curvature is given by
\begin{equation}
R = {G^{mn}R_{nm}}+2{G^{\eta\xi}R_{\xi\eta}}
 = R^{(g)}(1-if\xi\eta)-6g/\Lambda^2.
\end{equation}
Since $\sqrt{-G..}= -i\Lambda^2\sqrt{-g..}(1+2if\xi\eta)(1+ig\xi\eta)$, we 
obtain an action
\[
 I \equiv \frac{1}{2\Lambda^4}\int\,R\,\sqrt{G..}\,d^4x d\eta d\xi
 = \frac{1}{2\kappa^2}\int\, d^4 x\sqrt{-g..}
 \left[R^{(g)} +\lambda\right]
\]
where $\kappa^2 \equiv 8\pi G_N = \Lambda^2/(f+g)$, $R^{(g)}$ is the 
standard gravitational curvature and $\lambda = -6g(2f+g)/\Lambda^2(f+g)$ 
corresponds to a cosmological term. 

\medskip
\noindent
(2) Our second example leaves property space flat (in the $\eta,\xi$ sector)
but introduces a U(1) gauge field $A$, governed by the metric,
\[
 \left( \begin{array}{ccc}
 G_{mn} & G_{m\xi} & G_{m\eta}\\
 G_{\xi n} & G_{\xi\xi} & G_{\xi\eta}\\
 G_{\eta n} & G_{\eta\xi} & G_{\eta\eta}
          \end{array} \right)\!\!  =\!\!
 \left( \begin{array}{ccc}
   g_{mn}(1\!+\!if\xi\eta)\!+\!2i\Lambda^2\xi A_mA_n\eta & 
                     i\Lambda^2A_m\xi & i\Lambda^2A_m\eta \\
           i\Lambda^2 A_n\xi  & 0 & -i\Lambda^2\\
           i\Lambda^2A_n\eta & i\Lambda^2 & 0
          \end{array} \right).
\]
Simplify the analysis somewhat by going to flat (Minkowski) space first 
as there are then fewer connections. After some work ($F_{mn}\equiv A_{m,n}
-A_{n,m}$) one obtains,
\begin{eqnarray*}
{\Gamma_{\xi\eta}}^\xi&=&
 {\Gamma_{\xi\eta}}^\eta ={\Gamma_{\xi\eta}}^k = 0,\\
{{\Gamma_{m\xi}}^\xi} &=& {{\Gamma_{m\eta}}^\eta}
 =i\Lambda^2A^lF_{lm}\xi\eta/2,
\quad {{\Gamma_{m\xi}}^\eta}=-{{\Gamma_{m\eta}}^\xi}
  = A_m,\\
{{\Gamma_{m\xi}}^l}&=& i\Lambda^2{F^l}_m\xi/2,\quad 
   {{\Gamma_{m\eta}}^l} = i\Lambda^2{F^l}_m\eta/2,\\
{{\Gamma_{mn}}^\xi} &=& -A_mA_n\xi -(A_{m,n}+A_{n,m})\eta/2,\\
{{\Gamma_{mn}}^\eta} &=& -A_mA_n\eta +(A_{m,n}+A_{n,m})\xi/2,\\
{{\Gamma_{mn}}^k} &=&  i\Lambda^2(A_m{F^k}_n+A_n{F^k}_m)\xi\eta.
\end{eqnarray*}
Other Christoffel symbols can be deduced through symmetry of indices. Hence
\[
{R_{km}}={{R^l}_{klm}-{R^\xi}_{k\xi m}-{R^\eta}_{k\eta m}}=
 -i\Lambda^2(A_{k,l}+A_{l,k}){F^l}_m\xi\eta/2 + {\rm total~der.}
\]
\[
{R_{k\xi}}={R^l}_{kl\xi}+{R^\xi}_{k\xi\xi}+{R^\eta}_{k\eta\xi}= 
i\Lambda^2[{F^l}_{k,l}\xi/2+A^lF_{k,l}\eta]+{\rm total~der.}
\]
\[
{R_{k\eta}}={R^l}_{kl\eta}+{R^\xi}_{k\xi\eta}
  +{R^\eta}_{k\eta\eta} = 
i\Lambda^2[{F^l}_{k,l}\eta/2-A^lF_{k,l}\xi]+{\rm total~der.}
\]
\[
{R_{\xi\eta}}=-\Lambda^4F_{kl}F^{lk}\xi\eta.
\]
Then covariantize by including the gravitational component 
$g_{mn}(1+if\xi\eta)$ to end up with the total curvature:
\begin{eqnarray*}
R &=& {G^{mn}R_{nm}} + 2{G^{m\xi}R_{\xi m}} + 
  2{G^{m\eta}R_{\eta m}}+2{G^{\eta\xi}R_{\xi\eta}}\\
 &\rightarrow& R^{(g)} -3i\Lambda^2 g^{km}g^{ln}F_{kl}F_{nm}\xi\eta/2
\end{eqnarray*}
Finally rescale $A$ to identify the answer as electromagnetism + gravitation: 
\[
\int R\sqrt{-G..}d^4x d\eta d\xi /4\Lambda^4 =
  \int d^4x \sqrt{-g..} \left[R^{(g)}/2\kappa^2 - F^{kl}F_{kl}/4\right],
\]
where $\kappa^2=\Lambda^2/f =8\pi G_N$. It is a nice feature of the formalism 
that the gauge field Lagrangian arises from space-property terms --- like
the standard K-K model (from the tie-up between ordinary space-time 
and the fifth dimension).

Where do we go from these two examples? Well some generalizations come to mind:
\begin{itemize}
 \item replace property couplings $f$ and $g$ by two fields (dilaton and
  Higgs),
 \item combine the two models; this should lead to gravity + em + cosmic const.,
 \item extend fully to five $\zeta$; it is easy enough to incorporate the standard
 gauge model and one can even entertain a GUT $SU(5)$ of some ilk,
 \item work out the particle mass spectrum from all the $\langle\phi\rangle$.
\end{itemize}

\section*{Acknowledgements}
A considerable amount of the material has been drawn from previous papers written
with several collaborators including Peter Jarvis, Ruibin Zhang, Roland
Warner and Martin White. I would like to take the opportunity to thank them all 
for their valuable insights and help.


\newpage
\begin{thebibliography}{9}
\bibitem{RD1} R.~Delbourgo, {\em J. Phys. A} {\bf 39}, 5175 (2006).
\bibitem{RD2} R.~Delbourgo, {\em J. Phys. A} {\bf 39}, 14735 (2006). 
\bibitem{Neg} A.~McKane, Phys. Lett. {\em A76}, 22 (1980); \\
 P.~Cvitanovic and A.D.~Kennedy, Phys. Scripta {\bf 26}, 5 (1982);\\
 I.G.~Halliday and R.M.~Ricotta, Phys. Lett. {\bf B193}, 241 (1987);\\
 G.V.~Dunne, J. Phys. {\bf A22}, 1719 (1989).
\bibitem{Groups} P.D.~Jarvis and M.~White, Phys. Rev. {\bf D43}, 4121 (1991);\\
 R.~Delbourgo, P.D.~Jarvis, R.C.~Warner, Aust. J. Phys. {\bf 44}, 135 (1991).
\bibitem{KKF} R.~Delbourgo, S.~Twisk and R.~Zhang, Mod. Phys. Lett. {\bf A3}, 
  1073 (1988);\\
 P.~Ellicott and D.J.~Toms, Class. Quant. Grav. {\bf 6}, 1033 (1989);\\
 R.~Delbourgo and M.~White, Mod. Phys. Lett. {\bf A5}, 355 (1990).
\end{thebibliography}
\end{document}